\documentclass{cs20proc}
\def\ref{\par\noindent\hangindent=0.6in\hangafter=1}
\editors{S. J. Wolk}
\publisher{Zenodo}
\conference{The 20th Cambridge Workshop on Cool Stars, Stellar Systems, and the Sun}
\conferencedate{2018}

\title{High Resolution Near-Infrared Spectroscopy of Cool Dwarf Stars}
\author{Andrea Dupree$^{1}$, Nancy Brickhouse$^{1}$, Jonathan Irwin$^{1}$,
Robert Kurucz$^{1}$, and Elisabeth Newton$^{2}$}

\affiliation{$^{1}$ Center for Astrophysics | Harvard \& Smithsonian, 
Cambridge, MA USA\\
$^{2}$ Massachusetts Institute of Technology, Cambridge, MA USA and Dartmouth 
College, Hanover, NH USA}

\shorttitle{Near IR Spectra}
\shortauthors{Dupree et al.}

\abs{We present results from a near infrared survey
of the He I line ($\lambda$10830\AA)  in cool dwarf stars  
taken with the PHOENIX spectrograph at the
4-m Mayall telescope at Kitt Peak National Observatory. Spectral synthesis
of this region reproduces some but not all atomic and molecular features.  
The equivalent width of the He line appears directly correlated with  
the soft X-ray stellar surface
flux except among the coolest M dwarf stars, where the helium is
surprisingly weak.}  

\begin{document}

\maketitle

\section{Introduction}
The near IR spectra of cool stars are little studied to date, however they
contain diagnostics useful to identify warm plasma via the He I transition 
at $\lambda$10830\AA\  marking the presence of $\sim$15000K material.  In addition
the line profile can indicate mass motions because the transition arises
from a metastable level, giving it a long lifetime to allow tracking
of atmospheric dynamics.  This is very pronounced in luminous
cool stars (Dupree et al. 2009). Most recently, the 
helium line has emerged as a significant  
tracer of escaping exoplanet atmospheres (Oklop\u{c}i\`{c}, A. \& 
Hirata, C. M. 2018; Spake et al. 2018).

\begin{figure}[h!]
\includegraphics[scale=0.42]{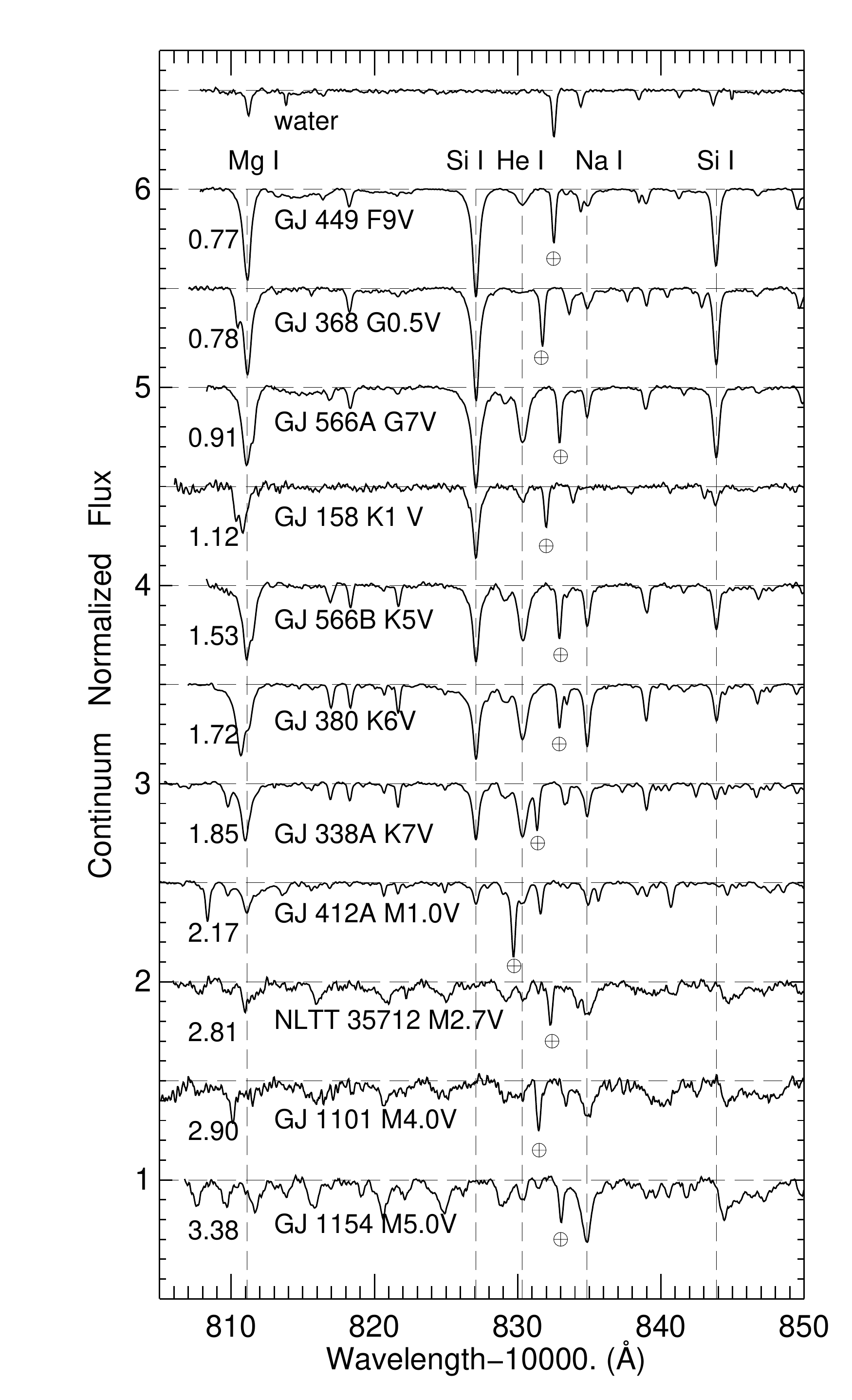}
\caption{Near-IR spectra of cool dwarf stars.
GAIA2 colors ({\it BP $-$ RP}) are marked (Evans et al. 2018).
Note the weakening of most atomic lines with later spectral
type and the increasing strength of the Na I transition. The He I absorption
exhibits varying strength discussed in Section 3.}
\end{figure}
\begin{figure*}[!ht]
\centering
\includegraphics[scale=0.5]{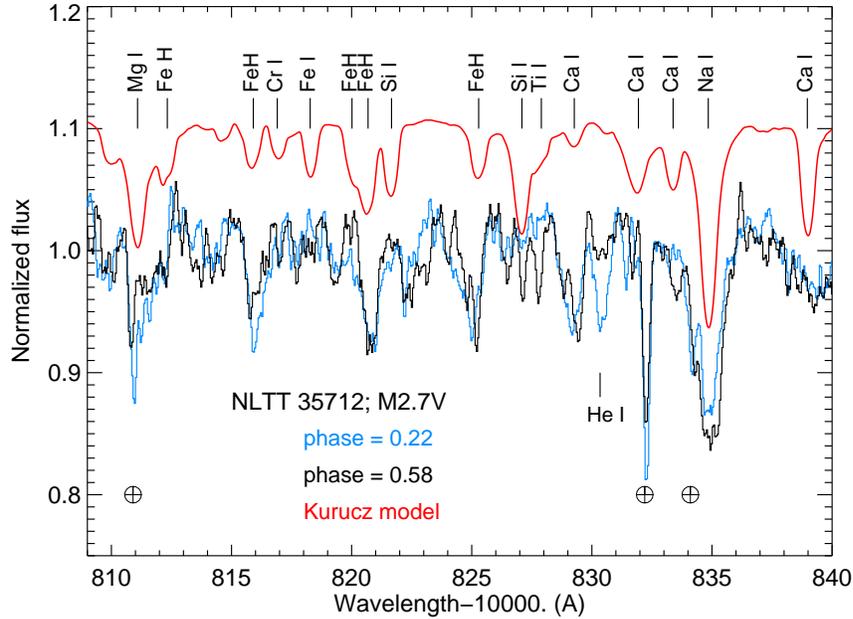}

\caption{PHOENIX near-IR spectrum taken at the 4-m Mayall Telescope at KPNO.
Spectrum calculated with a Kurucz LTE model  is shown. The strong
atomic features are reproduced, however several strong features remain
unidentified.}

\end{figure*}

\section{Near-IR Spectra of Cool Dwarfs}
The PHOENIX spectrograph at the 4-m Mayall telescope at KPNO was
used during February 2013 and 2014 to obtain near-IR spectra with  the 
J9232 filter. A summary of the exposures is given in Table 1. The 
section near $\lambda$10830 is shown in Fig. 1 
where the stars are ordered according to their photometric colors 
as measured by GAIA DR2 (Evans et al. 2018).  

\vspace*{0.2in}
\begin{tabular}{llr}
\multicolumn{3}{l}{Table 1: Phoenix Spectral Observations}\\ \hline\hline
Target & Date & Exposure$^1$\\ \hline
       &      & s(repeat)\\ \hline\hline
GJ 158&2456703.6218&600(2)\\
GJ 338A&2456704.7003&900(2)\\
GJ 368&2456701.0066&60(4)\\
GJ 380&2456703.7250&600(2)\\
GJ 412A&2456703.0414&900(2)\\
GJ 449&2456701.0131&60(4)\\
GJ 566A&2456702.9759&60(2)\\
GJ 566B&2456702.9790&90(2)\\
GJ 1101&2456346.6375&1500(4)\\
GJ1154&2456346.8192&1500(4)\\
NLTT 35712&2456346.9779&1500(4)\\ \hline
\end{tabular}
\parbox[b]{2.5in}{$^1$Exposure times are listed; in the infrared spectral
region, exposures
on the target are paired with a spatial offset. The number of
paired exposures is shown in parentheses.}
\bigskip

\vspace*{0.2in}

The warmer objects, of 
spectral type F$-$K exhibit easily identifiable atomic transitions
of neutral species (i.e. Mg I, Si I, Na I) in addition to the
He I transition.  These atomic species become weaker and the
spectrum more complex in the cooler dwarf stars $-$ beginning with
late K spectral type and continuing into the  M dwarfs. The Na I
transition (10834.848\AA) appears as an anomaly, becoming stronger
through the spectra of the M dwarfs

To identify spectral features, an LTE photospheric model corresponding 
to the parameters of a $\sim$M2 V star was constructed 
(kurucz.harvard.edu) assuming  T$_{eff}$ = 3500K, log g=5, 
v$_{rot}$=10 km s$^{-1}$, and solar abundances. Several
molecules are also included: FeH, TiH, TiO, VO, and H$_2$O. The synthesized
spectrum is shown in Figure 2 along with  observed spectra of 
NLTT 35712. The dominant Na I feature is reproduced in the calculation, as
well as Mg I.  The He I transition does not appear in the synthesized spectrum
because the stellar model does not include a chromosphere.  
While there is good agreement in some features, many others are not
identified.  So there is room for improvement.

\bigskip

Particularly noteworthy in the spectrum of NLTT 35712 is the 
change in the strength of the He I line
at 10830\AA\ with stellar rotational phase.  The spectra (Fig. 2) were taken at two different
photometric phases of NLTT 35712: 0.22 (close to photometric minimum) and 0.58 (approaching
photometric maximum). The He I line is stronger near photometric minimum. 
A simple interpretation of the photometry based on solar behavior 
suggests that dark starspots 
are present near photometric minimum. If similar to  the solar example, 
these starspots result from magnetic activity and would be associated 
with enhanced heating and EUV and X-ray radiation.  High energy 
radiation can cause
an increase in strength of the of the helium transition resulting
from photoionization of He I and recombination into the metastable
triplet level, although this effect may be mitigated when high densities
are present (Sanz-Forcada and Dupree 2008).

\begin{figure*}[!ht]
\includegraphics[scale=0.37]{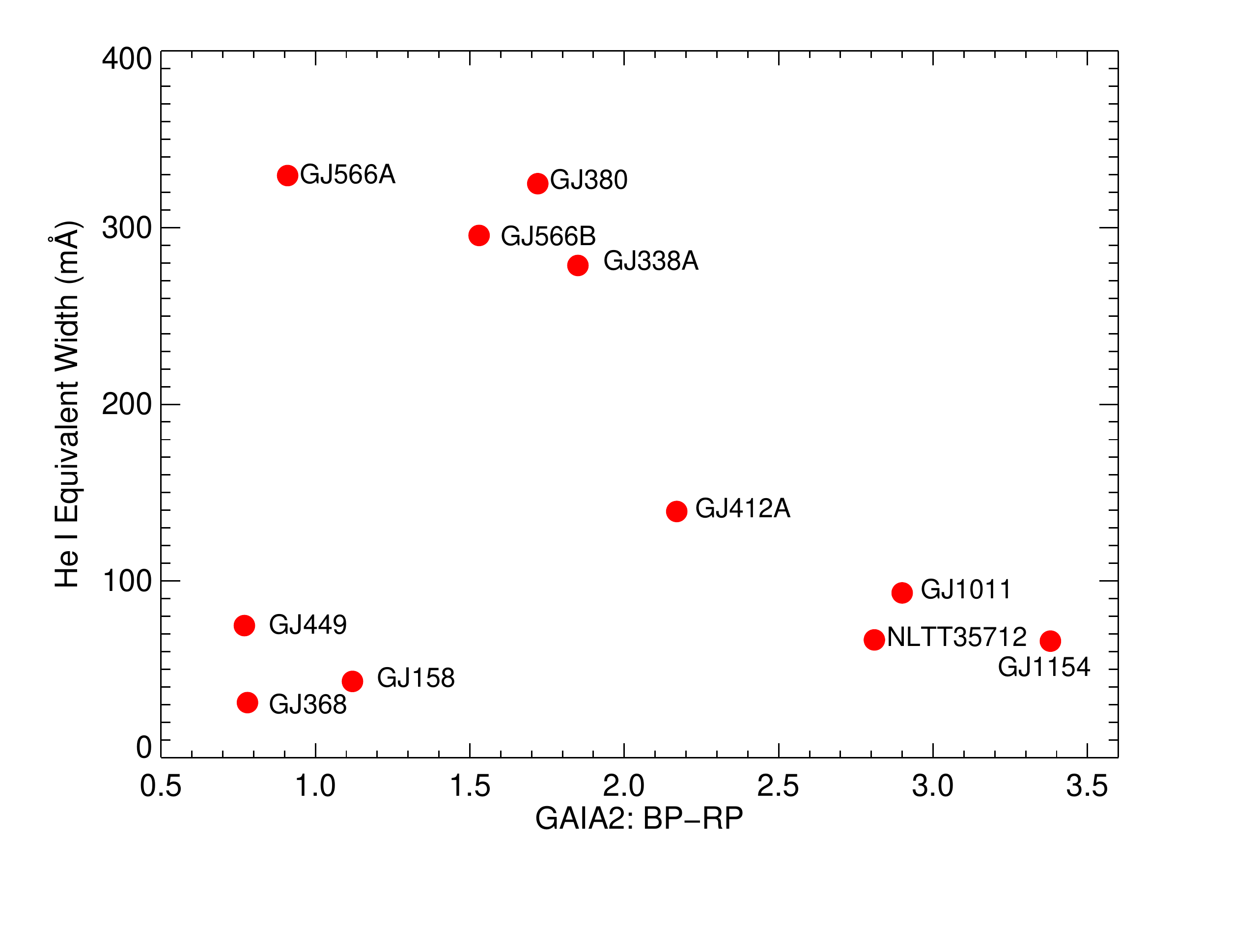}

\vspace*{-2.9 in}

\hspace*{+3.5in}
\includegraphics[scale=0.37]{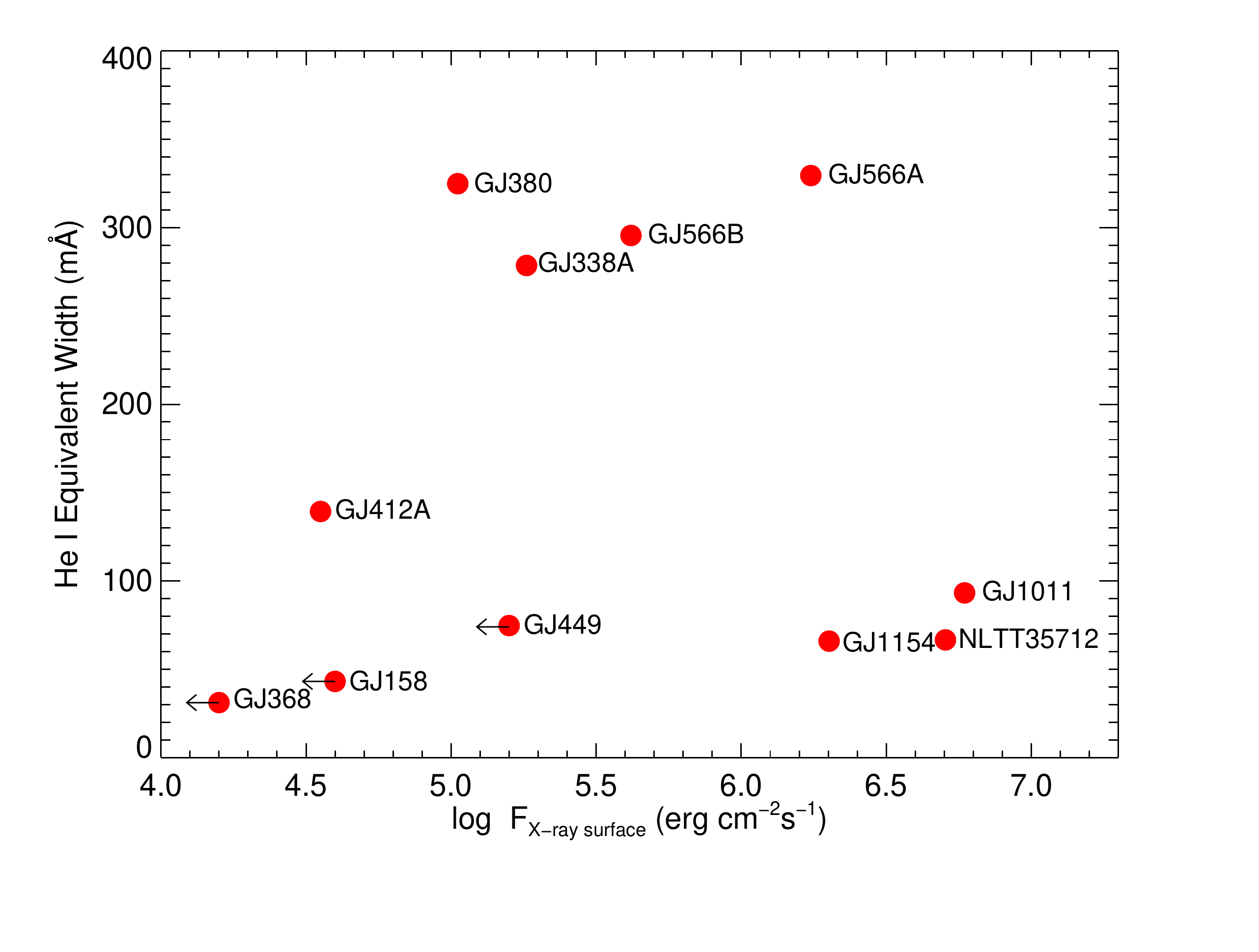}
\caption{{\it Left panel:} He I equivalent width as a function of GAIA  
color, {\it Right Panel:} He I equivalent width as a function of X-ray surface flux.}
\end{figure*}

\section{Helium Line Strength Correlations}

The equivalent width of the He I line is shown in Figure 3 as a function 
of the GAIA2 color ({\it left panel}) and the soft X-ray stellar surface flux ({\it 
right panel}). With the exception of the F through early K stars, the helium
equivalent width decreases with later spectral type.   The 3 early type stars  with  small
equivalent widths (GJ 368, GJ 449, and GJ 158) are single (Fuhrmann et al. 2017),
and in one case, metal poor (GJ 158 = HD 25329; [Fe/H]=$-$1.7; Gratton et al. 2000). 
The strength of the helium line does not depend on metallicity in dwarf 
stars (Takeda \& Takada-Hidai 2011; Smith et al. 2012), and the upper limits
are comparable among the 3 warmest stars (Pizzolato et al. 2000; Smith et al. 2016).   
X-ray surface fluxes,
considered to be more representative of magnetic activity (Johnstone and Gu\"del 2017),
are given in Table 2.  The 3 early-type stars
appear to illustrate the case  where the lower level of the helium line is populated
by collisions, and photoexcitation does not play a role.

At the same GAIA2 colors as the weak group, ($B-P \sim 1.5$) the stars GJ 566A and GJ 566B, 
although only slightly metal poor ([Fe/H]=$-$0.26, Allende Prieto et al. 2004),
possess stronger X-ray surface fluxes, by a factor of 10 or more than found
in GJ 367, GJ449, and GJ 158.  Also the
component A (GJ 566A) with the larger helium equivalent width  possesses an X-ray
surface flux, larger by a factor of 4 than GJ 566B.  Figure 3 suggests that the 
strength of the He I line is related to the  surface
X-ray flux for the F to early M  spectral types. 

However this correlation
changes dramatically in the coolest objects (GJ 1011, GJ 1154, and NLTT 35712) 
which have the largest surface X-ray flux
in the sample, yet display  weak helium absorption.  The cause of this change 
is not clear.  These 3 stars have 
strong H$\alpha$ emission (Newton et al. 2017) indicating a chromosphere
is present. These stars lie in the
fully convective regime where Morin et al. (2010) suggest that radically different
magnetic topologies can occur as compared to the more massive dwarf stars. The star GJ1154
has a very strong large-scale magnetic field that is mostly poloidal and 
axisymmetric.

The weakness of the helium line could suggest a distinctly different atmospheric structure.
Perhaps the chromosphere is substantially thinner, or smaller in 
surface extent  in the later type M dwarfs
as compared to the earlier types.   Without detailed models, it is difficult
to predict the structure. Inspection of the near-IR spectra in Fig. 1, shows
that the coolest stars have distinctly strengthened, Na I at 10834.9\AA\ as compared to the Si I
transition at 10827\AA. The Si I line arises from a 4.9eV level versus 3.6eV for
the Na I transition, suggesting the increasing presence of  cooler atmospheric
regions. 

\begin{table}
\bigskip
\begin{tabular}{lccrllccl}
\multicolumn{5}{l}{Table 2: Target Characteristics}\\ \hline\hline
Target & Gaia2 color  & Sp. Type&He I  &log F$_X^1$&Ref.  \\ \hline
       & {\it BP$-$RP}&         &m\AA\ & & \\ \hline\hline
GJ 158    &1.12&K1V  & 43  & <4.6  & 2,4  \\
GJ 338A   &1.85&K7V  & 279 & 5.26  & 2,5   \\
GJ 368    &0.78&G0.5V&31   & <4.2  & 2,3   \\
GJ 380    &1.72&K6V  &325  &  5.02 & 1,2   \\
GJ 412A   &2.17&M1.0V&139  &  4.55 & 1,2   \\
GJ 449    &0.77& F9V &75   & <5.2  & 2,3   \\
GJ 566A   &0.91&G7V  &330  &  6.24 & 2,6   \\
GJ 566B   &1.53& K5V &296  &  5.62 & 2,6   \\
GJ 1101   &2.90&M4.0V&93   &  6.77 & 1,2   \\
GJ 1154   &3.38&M5.0V&66   &  6.27 &1,2 \\
NLTT 35712&2.81&M2.7V&67   &  6.70 &2,7\\\hline

\end{tabular}
\begin{raggedright}
NOTE: $^1$X-ray stellar surface flux (erg cm$^{-2}$s$^{-1}$).\\
{\it References:} (1)Schmitt \& Lifke (2004);\\
NeXXus2 (http://www.hs.uni-hamburg.de/DE/For/Gal/Xgroup/nexxus/index.html)\\ 
(2) GAIA2, Evans et al. (2018)  (3) Pizzolato et al. (2000)\\  
(4) Smith et al. (2016) (5) Wood et al. (2012)\\  
(6) Johnstone \& G\"udel (2015) (7) Voges et al. (1999)
\end{raggedright}
\bigskip

\end{table}

On the other hand, studies of the near-ir helium line in metal-poor dwarf stars
suggest that a span in equivalent width values of He I of a factor amounting to 
2.5 to 3.5 can occur in stars
with similar X-ray fluxes (Takeda \& Takada-Hadai 2011). Additional stars
are needed to determine the extent of the variability here.

This work has made use of data from the European Space Agency (ESA) mission
{\it Gaia} ({\it https://www.cosmos.esa.int/gaia}), processed by the {\it Gaia}
Data Processing and Analysis Consortium (DPAC,
{\it https://www.cosmos.esa.int/web/gaia/dpac/consortium}). Funding for the DPAC
has been provided by national institutions, in particular the institutions
participating in the {\it Gaia} Multilateral Agreement.  This research has made use
of NASA's Astrophysics Data System Bibliographic Services and the VizieR catalogue access 
tool, CDS, Strasbourg, France.

\section{References}
\ref Allende Prieto, C., Barklem, P. S.,  Lambert, D. L., \& Cunha, K. 2004, A\&A, 420, 183
\ref Dupree, A. K., Smith, G. H., \& Strader, J. 2009, AJ, 138, 1485
\ref Evans, D. W., Riello, M., DeAngeli, F. et al. 2018, A\&A, 616, A4
\ref Fuhrmann, K., Chini, R., Kaderhandt, L., and Chen. Z. 2017, ApJ, 836, 139
\ref Gratton, R. G., Sneden, C., Carretta, E., \& Bragaglia, A. 2000, A\&A, 354, 169
\ref Johnstone, C. P., \& G\"udel, M.  2015, A\&A, 578, A129
\ref Morin, J., Donati, J.-F., Petit, P. et al. 2010, MNRAS, 407, 2269 
\ref Newton, E. R., Irwin, J., Charbonneau, D. et al. 2017, ApJ, 834, 85
\ref Oklop\u{c}i\'{c}, A. \& Hirata, C. M. 2018, ApJ, 855, L11
\ref Pizzolato, N., Maggio, A., \& Sciortino, S. 2000, A\&A, 361, 614 
\ref Sanz-Forcada, J., \& Dupree, A. K. 2008, A\&A, 488, 715
\ref Schmitt, J. H. M. M. \& Liefke, C. 2004, A\&A,  417, 651
\ref Smith, G. H., Dupree, A. K., \& G\"unther, H. M. 2016, AJ, 152, 43
\ref Smith, G. H., Dupree, A. K., \& Strader, J. 2012, Pub. ASP, 124, 1252
\ref Spake, J. J., Sing, D. K., Evans. T. M., et al. 2018, Nature, 557, 68
\ref Takeda, Y., \& Takada-Hidai, M. 2011, PASJ, 63, S547
\ref Voges, W., Aschenbach, B., Boller, Th. et al. 1999, A\&A, 349, 389
\ref Wood, B. E., Laming, J. M. \& Karovska, M. 2012, ApJ, 753, 76
\end{document}